\documentclass{PoS}

\title{Axial-Current Matrix Elements in Light Nuclei from Lattice QCD}

\ShortTitle{Axial-Current Matrix Elements in Light Nuclei from Lattice QCD}

\author{
\speaker{Martin J. Savage}
        \thanks{Presenting on behalf of the NPLQCD collaboration~\cite{NPLQCD}.
        We
        thank the Kavli Institute for Theoretical Physics for hospitality during the writing of these proceedings,
        supported in part by the National Science Foundation under Grant No. PHY11-25915        
        },
         Emmanuel Chang and Michael L. Wagman \\
       Institute for Nuclear Theory, Seattle, Washington 98195-1550, USA.\\
       E-mail: \email{mjs5@uw.edu}}

\author{Silas R. Beane\\ 
Department of Physics,
	University of Washington, Box 351560, Seattle, WA 98195, USA.
 	 }

 \author{Zohreh Davoudi, William Detmold and Phiala E. Shanahan\\
 	Center for Theoretical Physics, 
 	Massachusetts Institute of Technology, 
 	Cambridge, MA 02139, USA.
	}

 \author{Kostas~Orginos\\
 Department of Physics, College of William and Mary, Williamsburg,
 	VA 23187-8795, USA and 
Jefferson Laboratory, 12000 Jefferson Avenue, 
 	Newport News, VA 23606, USA.
  	 }

 \author{Brian C. Tiburzi\\
 Department of Physics, The City College of New York, New York, NY 10031, USA and
 Graduate School and University Center, The City University of New York, New York, NY 10016, USA and 
 RIKEN BNL Research Center, Brookhaven National Laboratory, Upton, NY 11973, USA. 
 }

\author{Frank Winter\\
Jefferson Laboratory, 12000 Jefferson Avenue, 
	Newport News, VA 23606, USA.
 	 }

\abstract{
I present results from the first lattice QCD calculations of axial-current matrix elements in light nuclei, 
performed by the NPLQCD collaboration. 
Precision calculations of these matrix elements, and the subsequent extraction of multi-nucleon axial-current operators, 
are essential in refining theoretical predictions of the proton-proton fusion cross section,
neutrino-nucleus cross sections and $\beta\beta$-decay rates of nuclei. 
In addition, they are expected to shed light on the phenomenological quenching of $g_A$ that is required in 
nuclear many-body calculations.  
}

\FullConference{38th International Conference on High Energy Physics\\
  3-10 August 2016\\
Chicago, USA}

\def\hetppn{$^{3} {\rm He}$}
\def\htpnn{$^{3} {\rm H}$}

\begin{document}

\section{Introduction}
\noindent
Axial-current matrix elements play a central role in important aspects of nuclear and particle physics research.
For example, the axial-current matrix element in the nucleon is intimately related to the long-range components of the 
strong nuclear force mediated by pions.  In addition,
the matrix element between the deuteron and proton-proton continuum at low-energies 
is responsible for the proton-proton fusion process that initiates the solar-burning cycle.
A number of experiments that are currently in production, or are planned for the near-term, depend 
upon a  knowledge of these matrix elements with varying degrees of precision.
From a theoretical standpoint, some of the important quantities are easier to determine than others. 
Lattice QCD calculations have matured to the point where early calculations of the properties and interactions of 
light nuclei are now being performed, 
albeit with unphysical values of the light-quark masses and without the electromagnetic interaction.
Programs have been established that are on track to perform precision calculations of 
important low-energy nuclear quantities in the not-so-distant future.
Reproducing quantities, such as the excitation spectra of light nuclei and nucleon-nucleon scattering 
parameters, within the experimental uncertainties will provide verification of the essential elements of 
LQCD calculations.  
Analogous verification steps for the nuclear matrix elements of external probes, such as the 
electroweak interactions, are  required for a complete quantification of the uncertainties 
associated with any suite of LQCD calculations.
The spectra of the s-shell nuclei and hypernuclei have been calculated
at the SU(3)-symmetric point with a pion mass of $m_\pi\sim 805~{\rm MeV}$, 
along with the nucleon-nucleon scattering phase-shifts and parameters in the s-wave and also 
higher-partial waves.  Analogous calculations continue to  be performed at lighter quark masses, and 
calculations in the two-nucleon systems are being performed at the physical point. 

The objective of LQCD calculations in multi-nucleon systems is to provide precise and reliable 
calculations of an array of quantities in few-nucleon systems
in order to refine the nuclear forces 
and interactions that are employed in nuclear many-body calculations of more complex nuclei and systems.
In particular, LQCD calculations are starting to constrain multi-nucleon interactions and 
the correlated two-nucleon interactions with electroweak probes, that are  to be propagated into 
larger nuclei using effective field theory (EFT) techniques.
In the case of axial-current matrix elements, it is known that the correlated two-nucleon interactions 
with axial currents play an important role, and the leading low-energy effects are parameterized by the 
counterterm $L_{1A}$ in  pionless EFTs,  
and analogous counterterms in pionful EFTs.
In addition to the contributions from $g_A$, the proton-proton fusion process, tritium $\beta$-decay and 
neutrino-deuteron scattering and break-up all receive contributions from $L_{1A}$.  
Higher energy processes that lie outside the radii of convergence of the low-energy EFTs are more 
challenging from the hadronic standpoint and reliable theoretical tools for GeV-scale 
processes remain to be established.

\section{Status of Lattice QCD Calculations of Light Nuclei}
\begin{figure}[!ht]
  \centering
  \includegraphics[width=0.55\textwidth]{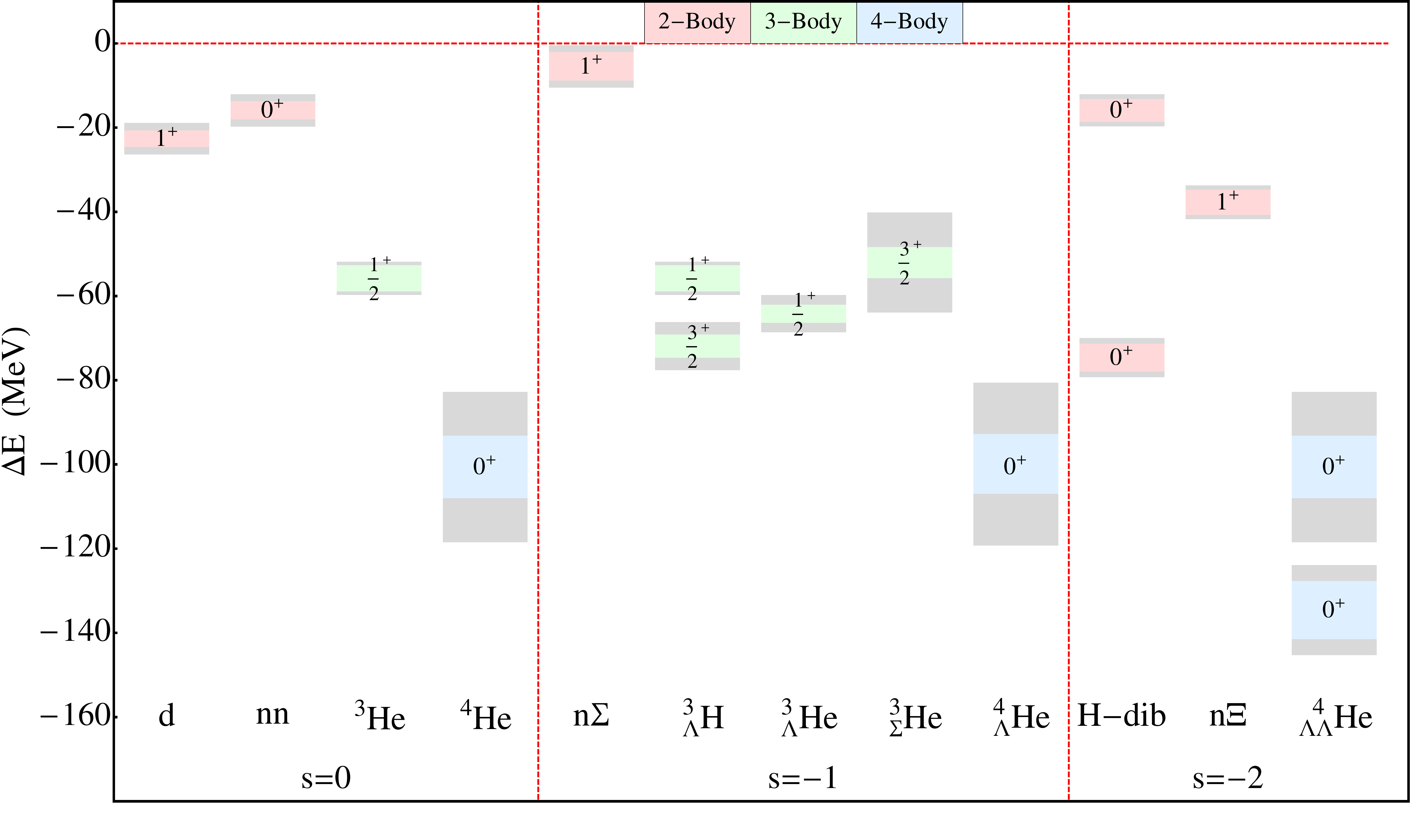}
  \caption{A compilation of the nuclear energy levels at $m_\pi\sim800$ MeV
with spin and parity  $J^\pi$ ~\cite{Beane:2012vq}.
}
  \label{fig:NuclearSummary}
\end{figure}

\begin{figure}[!ht]
  \centering
     \includegraphics[width=0.45\textwidth]{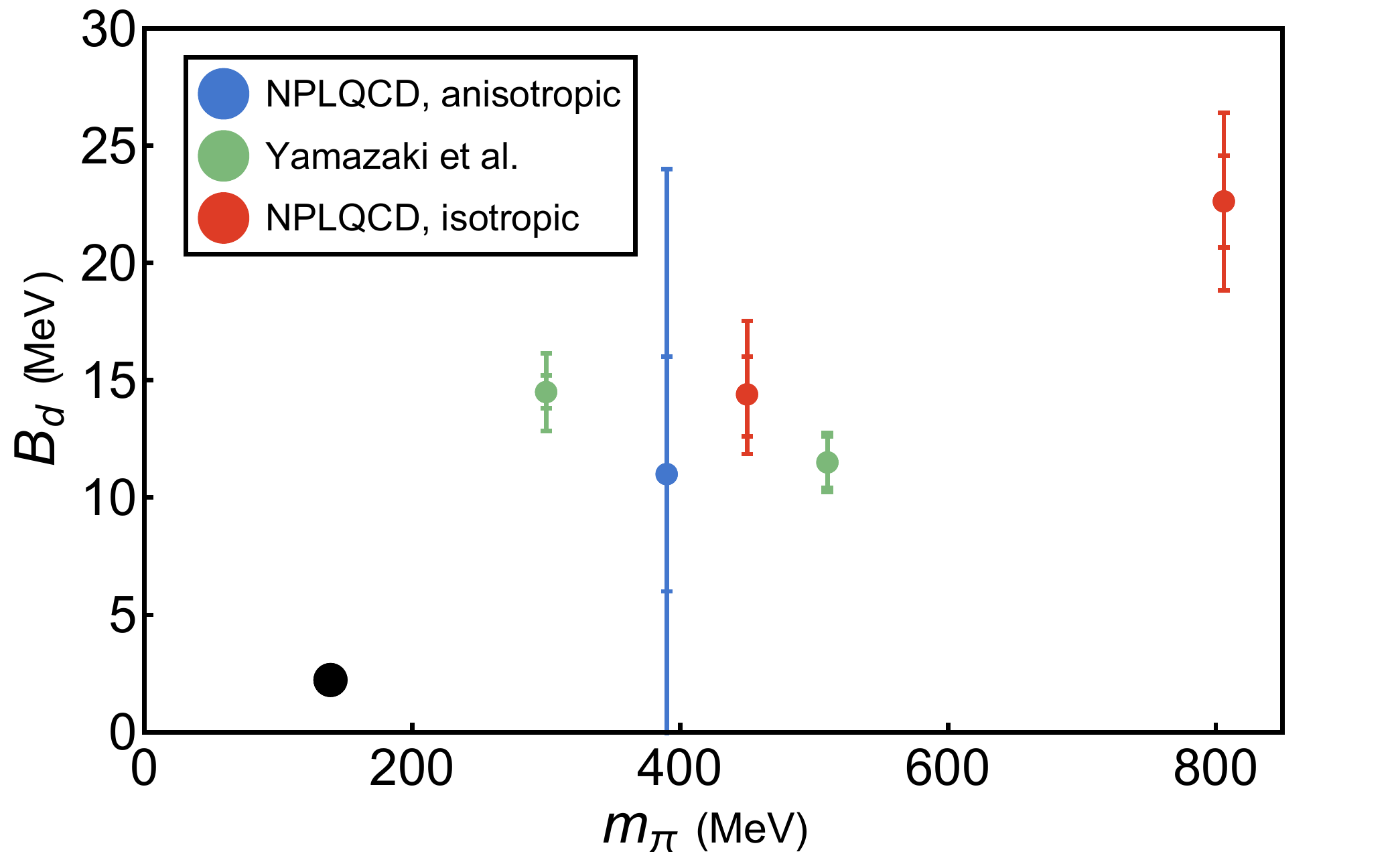}\ \ 
     \includegraphics[width=0.45\textwidth]{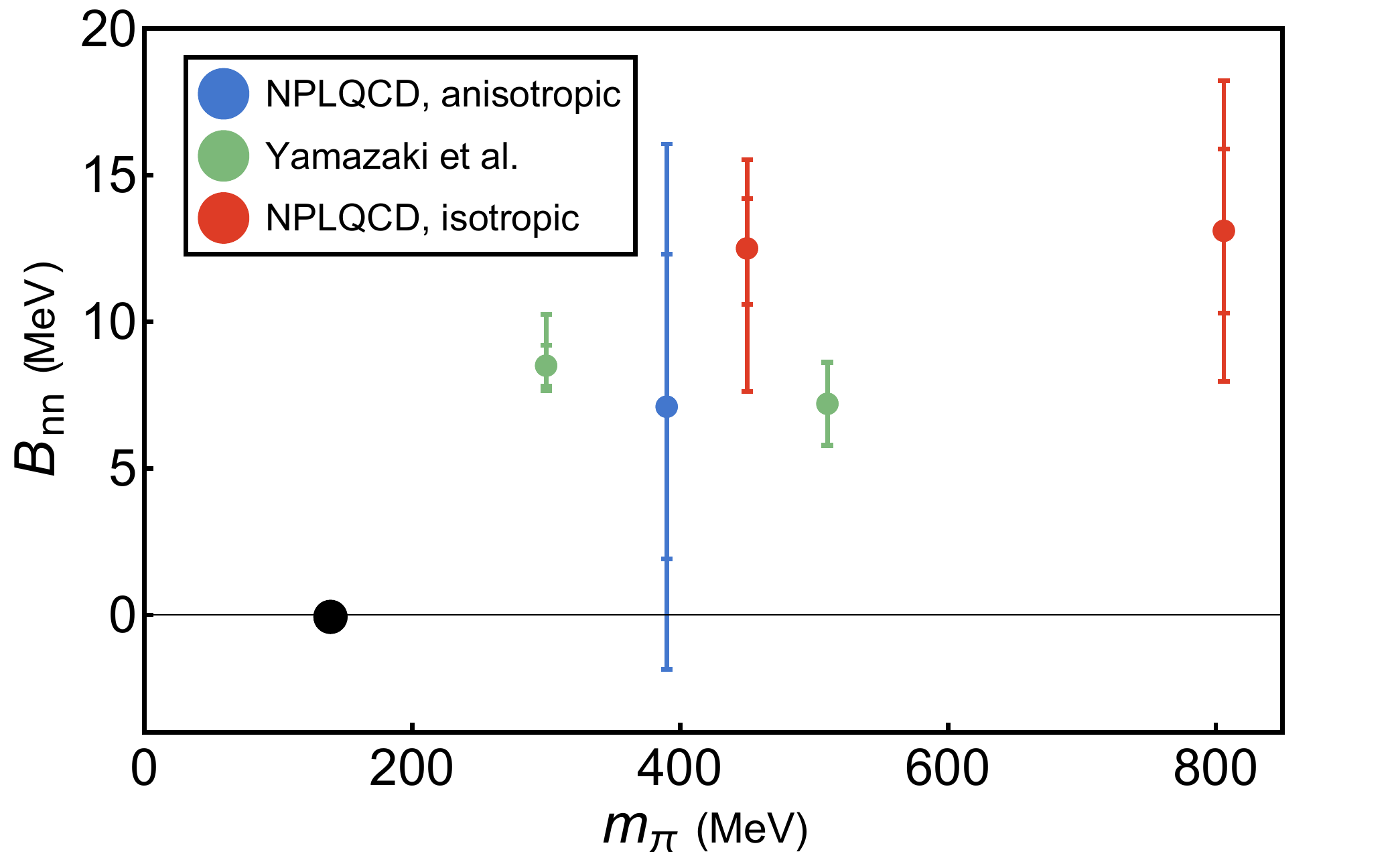}
     \caption{
     The deuteron (left panel),  and
       dineutron (right panel) binding energies~\protect\cite{Beane:2011iw,Beane:2012vq,Yamazaki:2015asa,Orginos:2015aya}.
       }
  \label{fig:Ball}
\end{figure}

\noindent
A summary of the nuclear and hypernuclear binding energies obtained by NPLQCD
at a pion mass of  $m_\pi\sim 805~{\rm MeV}$~\cite{Beane:2012vq} 
is shown in Fig.~\ref{fig:NuclearSummary},
from which it is seen that the  binding energy per baryon is  
larger than that found in nature.
Consistent a  large-$N_c$ analysis, 
the difference in binding between the deuteron and the dineutron is found to be
smaller than the splittings to the other states.
For systems comprised of three baryons, the
ground state of \hetppn\ (and its isospin partner \htpnn) are cleanly identified,
as are the ground states, and even one excited hypernucleus.
Attaining accurate binding energies for any given light-quark masses will require the inclusion of
electromagnetic and isospin-breaking effects.
A deeper understanding of the origin and phenomenological interpretation of the calculated binding energies 
requires a series of nuclear few-body calculations.
 In particular, it is important to understand the relative
contribution from the two-body, three-body, and higher-body
contributions to the $A\ge 3$ nuclei and hypernuclei, which can only be
accomplished using modern few-body techniques, as recently tackled 
with the pionless EFT
in Refs.~\cite{ Barnea:2013uqa,vanKolck:2015mra,Kirscher:2015tka}.
Using the $\alpha$-particle prediction for verification,
it was shown that three-nucleon forces are clearly present in the  LQCD results.
Figure~\ref{fig:Ball} shows the binding energy of  the deuteron and 
dineutron~\protect\cite{Beane:2011iw,Beane:2012vq,Yamazaki:2015asa}
obtained by the PACS collaboration and NPLQCD
as a function of the pion mass.~\footnote{
The HALQCD collaboration does not observe bound states in solutions to the Schr\"odinger equation with 
$U_E(r,r^\prime)\rightarrow V(r)$ extracted from
4-point Green functions produced with wall-sources.  For example, see Ref.~\cite{Aoki:2013tba,Iritani:2016jie}.
}
It is exciting to see nuclei emerge from QCD for a range of
the light-quark masses, and such calculations are crucial in dissecting and refining
the chiral nuclear forces. 
However,  calculations at even lighter pion masses remain critically important.

\section{Magnetic Interactions}
\begin{figure}[!ht]
  \centering
  \includegraphics[width=0.45\textwidth]{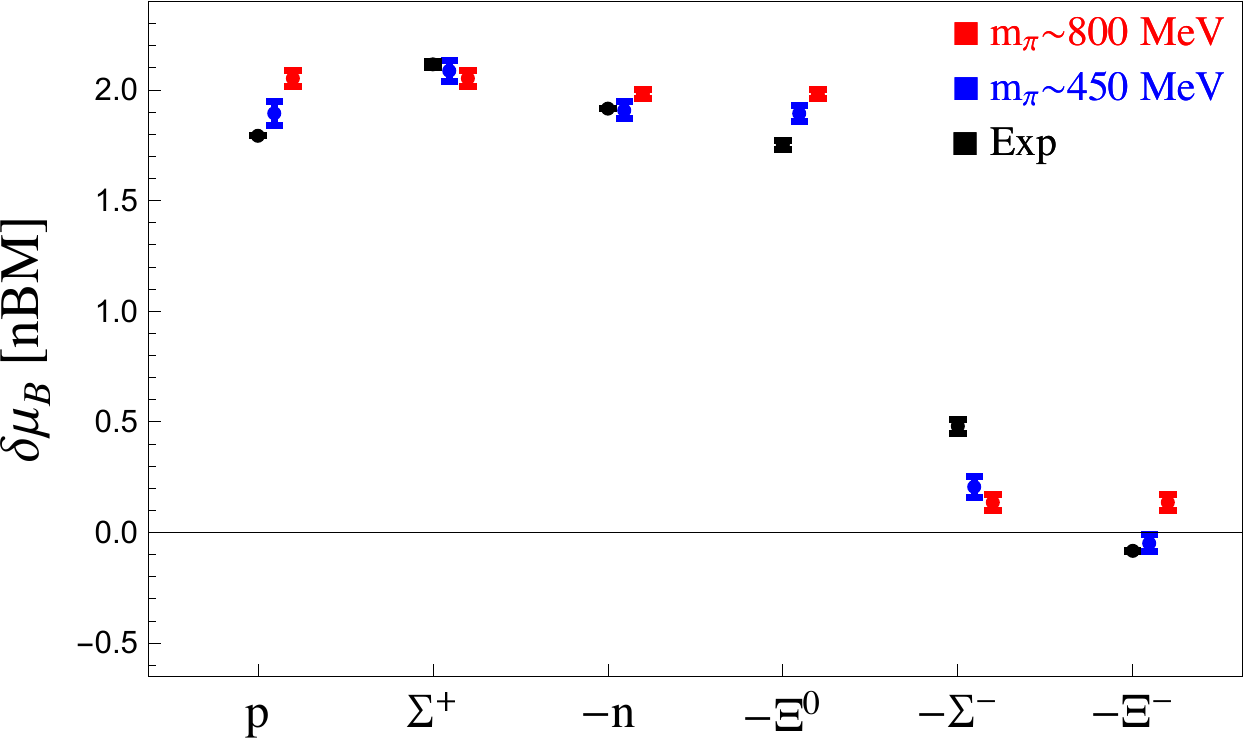}
  \includegraphics[width=0.5\textwidth]{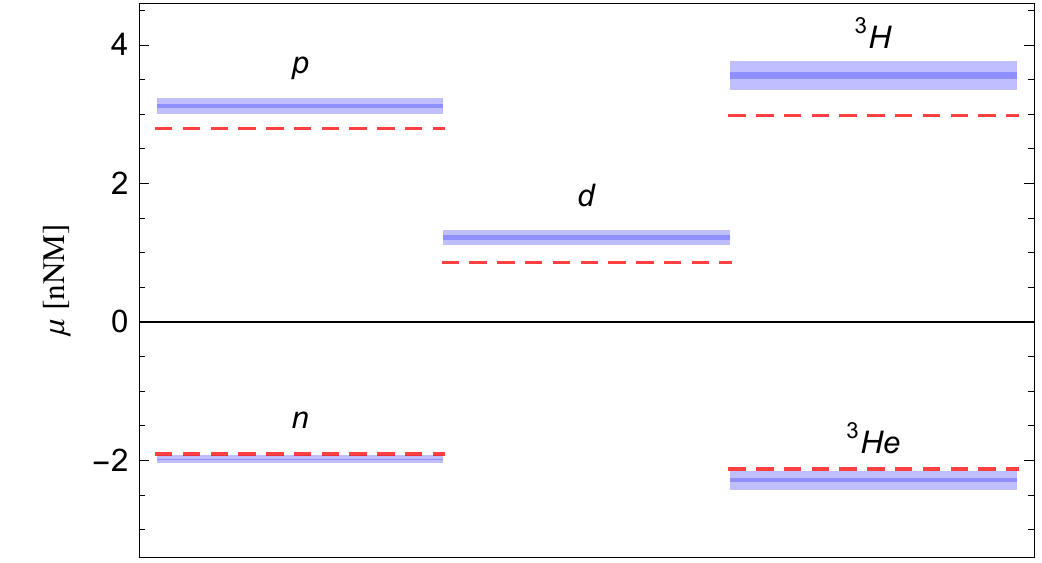}
  \caption{The anomalous magnetic moments of the lowest-lying octet of 
  baryons~\protect\cite{Parreno:2016fwu} (left panel) in units of natural Baryon Magnetons, 
  and the magnetic moments of the light nuclei in units of natural Nuclear Magnetons
  calculated at $m_\pi\sim 805~{\rm MeV}$ 
  compared with experiment~\protect\cite{ Beane:2014ora,Chang:2015qxa}  (right panel).
}
  \label{fig:anomMM}
\end{figure}

\noindent
The magnetic moments and polarizabilities of the light nuclei have been
calculated at pion  masses of $m_\pi\sim 805~{\rm MeV}$ and $\sim 450~{\rm MeV}$~\cite{ Beane:2014ora,Chang:2015qxa} using background field methods. 
In addition,
the low-energy cross section for the simplest radiative capture process, $np\rightarrow d\gamma$, has been 
calculated~\cite{Beane:2015yha}.
A number of interesting features and observations emerge from these calculations, 
despite the unphysical masses of the light quarks, and extrapolations to the physical point have been accomplished.
Perhaps the most interesting observation is that when the  moments are 
given in units defined by the baryon mass, 
they are relatively insensitive to the light-quark masses, with little variation observed between $m_\pi\sim 805~{\rm MeV}$  and the physical point, as shown in Figure~\ref{fig:anomMM}.
Further, the deviations from  predictions of the naive single-particle shell model at the heavier quark 
masses are consistent with the corresponding experimental values, within the uncertainties of the calculations.

The very low-energy radiative-capture process $np\rightarrow d\gamma$ is dominated by the M1 matrix element,
which receives most of its contribution from the nucleon isovector magnetic moment and from
correlated two-nucleon interactions with the magnetic field (meson-exchange currents).
By performing LQCD calculations at the heavier quark masses~\cite{Beane:2015yha}, 
the counterterm in the low-energy EFT describing $np\rightarrow d\gamma$ has been determined 
and extrapolated to the physical point.   
Like the magnetic moments themselves, it is found to be quite insensitive to the 
masses of the light quarks within the uncertainties of the calculations.  
Using the experimentally measured scattering parameters and the LQCD determined counterterm,  
the post-dicted $np\rightarrow d\gamma$ cross section is found to agree with experiment within the uncertainties.

\section{Axial-Current Matrix Elements}
\noindent
Important low-energy axial-current matrix elements are starting to be determined from LQCD calculations, 
using  methods analogous to those used to determine magnetic matrix elements.  
Axial background fields are used to generate light-quark propagators, which in turn are used to generate the 
relevant axial-current matrix elements.   
At present, calculations are performed at unphysical values of the light quark masses and extrapolated to the 
physical point with EFTs. Further, only one spacetime volume with one discretization is employed.
Once the appropriate renormalization factor that relates the lattice axial current to that in the continuum is employed, 
the value of $g_A$ in the proton is recovered at $m_\pi\sim 805~{\rm MeV}$ within the uncertainties of the calculations. 

\begin{figure}[!ht]
  \centering
  \includegraphics[width=0.4\textwidth]{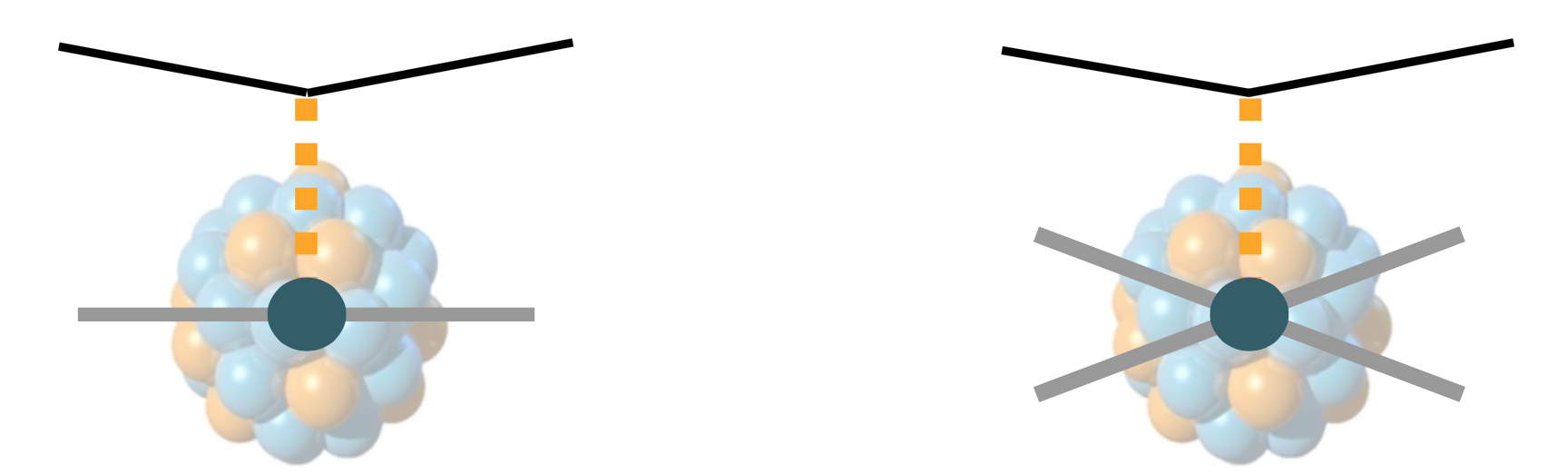}\ \ \ \ 
 \raisebox{-0.3\height}{   \includegraphics[width=0.4\textwidth]{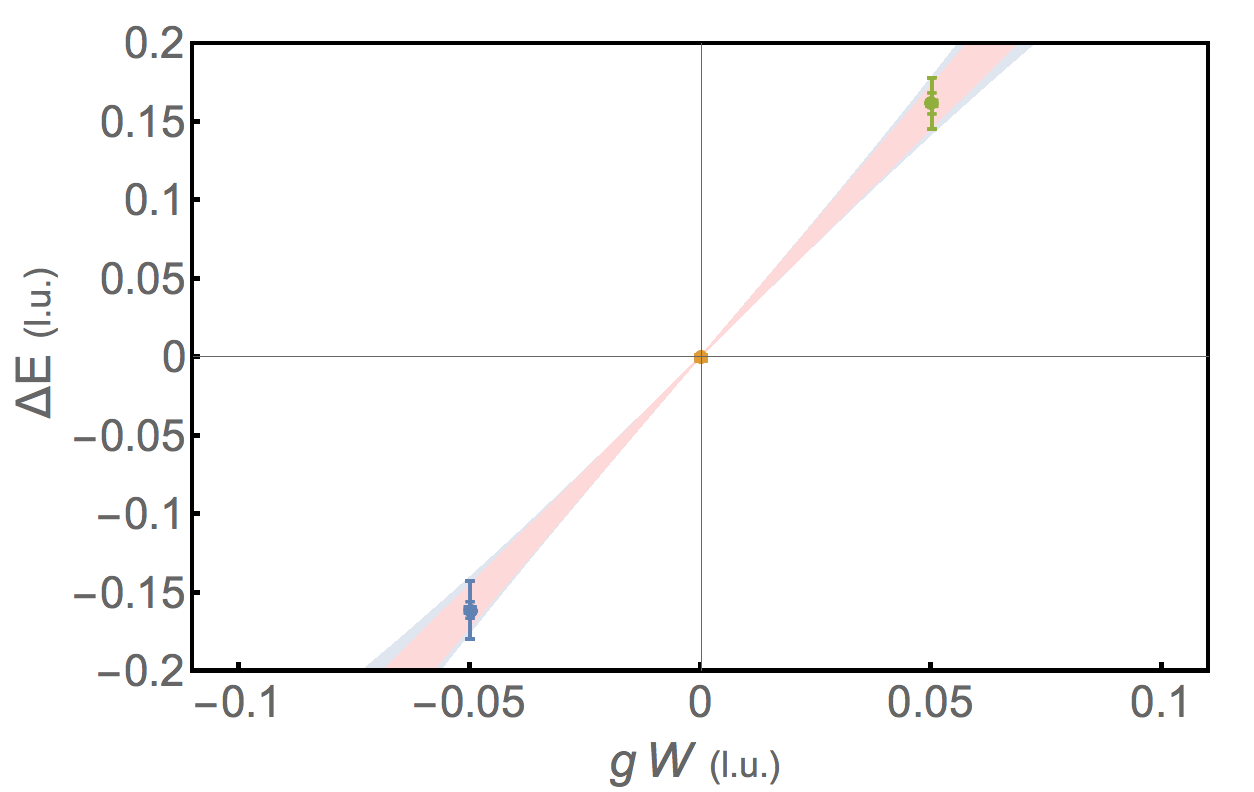}  }
  \caption{The left panel shows a cartoon of the single-nucleon and correlated two-nucleon axial-current interactions,
  while the right panel shows preliminary
  energy splittings between the two lowest-lying eigenstates in the $J_z=I_z=0$ two-nucleon sector as a 
  function of the background axial field.
}
  \label{fig:Nints}
\end{figure}

The leading matrix element contributing to low-energy proton-proton fusion, 
the nuclear reaction that initiates the solar-burning cycle, has  been calculated at $m_\pi\sim 805~{\rm MeV}$, leading to a constraint on the higher-order isovector counterterm in the pionless EFT, $L_{1A}$, as shown in Figure~\ref{fig:Nints}.  
As the low-energy operator structure for this process is the same as  for 
$np\rightarrow d\gamma$ and the matrix element in the nucleon, $g_A$, 
is also fairly insensitive to the light-quark masses over this interval,
a value and systematic uncertainty has been assigned at the physical point.  
Clearly, this is a first calculation of this process, and calculations over a range of pion masses 
and at the physical point are required.   
Figure~\ref{fig:Nints} shows the energy splittings between the two lowest-lying eigenstates 
in the $J_z=I_z=0$ two-nucleon sector as a  function of the background axial field, 
which can be used to constrain  $L_{1A}$.
The Gamow-Teller matrix element contributing to the $\beta$-decay of tritium can also be determined with 
axial background-field calculations.  
Such calculations  at $m_\pi\sim 805~{\rm MeV}$
lead to a value of $g_A(^3H)$ that is close to that of the proton, and consequently, by isospin symmetry, 
a value of the Gamow-Teller matrix element that is consistent with experiment.  
This is further indication of a weak pion mass dependence of the higher order  contributions in the low-energy EFT.

Finally,   the background-field technique enables a determination of the isotensor axial polarizability 
of the dinucleon spin-singlet system, which  contains the matrix element 
for $nn\rightarrow pp$ transitions relevant for $\beta\beta$-decay processes in nuclei.
The axial polarizability contributes to  energy shifts that depend  quadratically on the background axial field strength.

\section{Post-Presentation Progress}
\noindent
Our initial work on axial-current matrix elements was recently completed~\cite{Savage:2016kon}.
The Gamow-Teller matrix element contributing to tritium $\beta$-decay
was found to be 
${\rm GT} = 0.979(03)(10)$, which should be compared to the experimental value of 
${\rm GT}^{\rm expt} = 0.9511(13)$.
The S-factor for pp-fusion, in the absence of electromagnetism and isospin breaking, was found to be
$\Lambda(0) = 2.6585(06)(72)(25)$, assuming that the correlated two-nucleon interactions are independent of the light-quark masses and assigning a systematic uncertainty based upon what is observed in $np\rightarrow d\gamma$,
leading to a value of 
$L_{1A}=3.9(0.1)(1.0)(0.3)(0.9)~{\rm fm}^3$.
This is to be compared with the currently accepted value of $\Lambda(0) = 2.652(2)$.
These results are the first of such  calculations, and demonstrate that, with sufficient computing resources, 
axial-current matrix elements can be determined with precision from ongoing and future LQCD calculations.

\section{Summary and Outlook}
\noindent
Results of the first Lattice Quantum Chromodynamics calculations of 
matrix elements of the axial current in light nuclei have been presented.
In particular, first calculations of the Gamow-Teller matrix element contributing 
to the $\beta$-decay of tritium, and the low-energy cross-section for pp-fusion have been 
performed by the NPLQCD collaboration at a pion mass of $m_\pi\sim 805~{\rm MeV}$.
They yield results that are consistent with experiment and phenomenology, respectively, 
under the assumption of mild quark-mass dependence of the correlated two-nucleon contributions, 
consistent with what is found for $np\rightarrow d\gamma$.
Finally, 
this work paves the way for axial-current form factors of light nuclei to be addressed with LQCD.

%
%

\end{document}